\documentclass[aip,pof,amsmath,amssymb,reprint,floatfix]{revtex4-1}
\usepackage{graphicx,subfig,booktabs}

\usepackage[svgnames,table,hyperref]{xcolor} 
\definecolor{dblue}{rgb}{0 0.447 0.741}
\definecolor{dgreen}{rgb}{0.274 0.784 0.424}

\usepackage{dcolumn}
\usepackage{bm}

\usepackage[utf8]{inputenc}
\usepackage[T1]{fontenc}
\usepackage{mathptmx}
\usepackage{units}	
\usepackage{morefloats}

\newcommand{\bi}{ \boldsymbol}%

\usepackage[colorlinks,linkcolor=dblue,citecolor=dgreen,pagebackref]{hyperref}

\begin{document}


\title{Generating a tide-like flow in a cylindrical vessel by electromagnetic forcing}

\author{Peter J\"ustel}
  \altaffiliation{Helmholtz-Zentrum Dresden - Rossendorf, Bautzner
	Landstr. 400, 01328 Dresden, Germany}

\author{Sebastian R\"ohrborn}
  \altaffiliation{Helmholtz-Zentrum Dresden - Rossendorf, Bautzner
	Landstr. 400, 01328 Dresden, Germany}
	
\author{Peter Frick}
  \altaffiliation{Institute of Continuous Media Mechanics,1 Acad. Korolyov str., 614013 Perm, Russia}
  
 \author{Vladimir Galindo}
  \altaffiliation{Helmholtz-Zentrum Dresden - Rossendorf, Bautzner
    Landstr. 400, 01328 Dresden, Germany} 
    
 \author{Thomas  Gundrum}
  \altaffiliation{Helmholtz-Zentrum Dresden - Rossendorf, Bautzner
    Landstr. 400, 01328 Dresden, Germany}
    
 \author{Felix Schindler}
  \altaffiliation{Helmholtz-Zentrum Dresden - Rossendorf, Bautzner
	Landstr. 400, 01328 Dresden, Germany}

\author{Frank Stefani}
  \altaffiliation{Helmholtz-Zentrum Dresden - Rossendorf, Bautzner
	Landstr. 400, 01328 Dresden, Germany}

\author{Rodion Stepanov}
  \altaffiliation{Institute of Continuous Media Mechanics,1 Acad. Korolyov str., 614013 Perm, Russia}

\author{Tobias Vogt}
  \altaffiliation{Helmholtz-Zentrum Dresden - Rossendorf, Bautzner
    Landstr. 400, 01328 Dresden, Germany}

\date{\today}

\begin{abstract}
We show and compare numerical and experimental results on the electromagnetic 
generation of a tide-like flow structure in a cylindrical vessel which is 
filled with the eutectic liquid metal alloy GaInSn. Fields of various strengths 
and frequencies are applied to drive liquid metal flows. The impact of the 
field variations on amplitude and structure of the flows are investigated. 
The results represent the basis for a future Rayleigh-B\'enard experiment, 
in which a modulated tide-like flow perturbation is expected to  synchronize 
the typical sloshing mode of the large-scale circulation. A similar entrainment 
mechanism for the helicity in the Sun may be responsible for the 
synchronization of the solar dynamo with the alignment cycle of the tidally 
dominant planets Venus, Earth and Jupiter.

\end{abstract}



\maketitle


\section{Introduction} \label{introduction}
In a series of recent papers 
\cite{Weber2015,Stefani2016,Stefani2018,Stefani2019,Stefani2020}, an 
attempt was made to explain the remarkable empirical synchronization 
of the solar cycle with the 11.07 years alignment cycle of the tidally 
dominant planets Venus, Earth and Jupiter. The basic idea relies, 
first, on the tendency of the current-driven, kink-type ($m=1$) 
Tayler instability (TI) \cite{Tayler1973} to undergo helicity 
oscillations \cite{Weber2015} and, second, on the fact that these 
helicity oscillations can easily be entrained by tide-like forces 
with their typical $m=2$ azimuthal dependence \cite{Stefani2016}. 
While the tidal forces of the planets are indeed tiny, in this 
mechanism they are only needed as a catalyst to switch between 
left- and right-handed states of the TI, without, or just barely, 
changing its energy content. 

This theory is still highly speculative and further conceptual and 
numerical work will be needed for its verification. Additional 
experimental insight into this synchronization effect could also 
be helpful in this respect. Unfortunately, experiments on the very 
TI using liquid metals (which are, similarly as the plasma in the 
solar tachocline region, characterized by a low magnetic Prandtl 
number) have turned out especially complicated \cite{Seilmayer2012}, 
mainly due to the necessity to characterize the instability in a 
contactless manner in order not to generate any interfering 
electrovortex flow by the insertion of measurement probes such 
as ultrasonic transducers. 

Yet, the key idea of synchronizing the helicity, which is connected 
with an $m=1$ flow structure, 
by some tide-like $m=2$ perturbation
might have wider applicability than just for the TI case. 
An interesting candidate in this 
respect is the $m=1$ magneto-Rossby
wave at the  solar tachocline\cite{Dikpati2017,McIntosh2017,Tobias2017,Zaqarashvili2018}, 
which could 
open up a complementary route for synchronizing the solar dynamo 
by tidal forces. 

Large-scale circulation (LSC) 
\cite{Krishnamurti1981,Sano1989,Takeshita1996,Cioni1997,XiLamXia2004,BrownAhlers2006,Resagk2006,Xi2008}, 
which appears in 
Rayleigh-B\'enard convection (RBC) when thermal plumes erupt from the 
boundary layer and self-organize into a flywheel structure 
\cite{Kadanoff2001}, is
another candidate. Quite similarly as the TI,
the $m=1$ LSC also breaks spontaneously the axi-symmetry ($m=0$) 
of the underlying problem,
and becomes prone to secondary effects such as torsional \cite{Funfschilling2004}
and 
sloshing modes \cite{XiZhou2009,BrownAhlers2009},
reversals and even intermittent cessations \cite{BrownAhlers2006,Xi2008}.
Experiments on the  LSC problem were carried
out with different working fluids, including 
water \cite{Krishnamurti1981,BrownAhlers2006,Xi2008}, 
silicon oil \cite{Krishnamurti1981}, helium-gas \cite{Sano1989}, 
air \cite{Resagk2006}, 
liquid mercury \cite{Takeshita1996,Cioni1997},
liquid sodium \cite{Khalilov2018}, and the 
eutectic alloy GaInSn \cite{Wondrak2018,Zuerner2019}.

On closer consideration, the sloshing mode with its side-wise motion 
transverse to the primary LSC vortex, turns out to be connected with 
helicity oscillations. Some preliminary simulations have recently 
shown that this sloshing motion of the LSC can also be synchronized 
by an $m=2$ perturbation, although with a different frequency 
relation (2:1) than in the TI-case. This effect, which might have 
to do with the different numbers of vertically stacked vortices, 
needs further clarification. At any rate, the interaction of the 
sloshing mode of an LSC with some $m=2$ perturbation seems quite 
promising for experimentally evidencing the generic effect of 
helicity synchronization 
by tide-like forces.

There are different ways to experimentally realize such perturbations. 
In a zonal wind experiment, for example, a deformable wall was used 
to force a tide-like perturbation \cite{Morize2010}. When using a 
liquid metal, electromagnetic forcing, modulated in time, opens 
an alternative route for realizing an $m=2$ flow perturbation. 
Such a concept of tide-like forcing will be pursued in this paper, where
we focus, though, mainly on the spatial ($m=2$) aspect of the tidal 
flow, leaving
its time-dependence and its influence on the LSC to future studies. 

Our experiment utilizes the versatile MULTIMAG system \cite{Pal2009}, 
which allows for arbitrary superpositions of axial, rotating, and 
travelling magnetic fields. In a recent paper \cite{StepanovStefani2019}, 
various combinations of the six coils, which are normally used for 
producing the rotating magnetic field (RMF), have been evaluated with 
respect to their suitability to generate a typical tide-like $m=2$ 
flow perturbation. Although these preparatory simulations were made 
in the simple Stokes approximation, they were indeed useful to 
discriminate between different flow topologies when applying various 
combinations of the available coils. A very favourable flow structure 
appeared for the comparably simple configuration that two oppositely 
situated coils are fed with AC currents with a frequency of some tens of Hz.

Based on this prediction, we have carried out a corresponding experiment 
in which we measured the electromagnetically driven $m=2$ flow in a 
vessel filled with the eutectic alloy GaInSn. We will show that the 
measured flow structure corresponds well with the results of accompanying 
OpenFOAM$^\copyright$ simulations. This gives a solid basis for the 
combination of a LSC flow produced by thermal convection with a 
modulated $m=2$ force, which is planned for further experiments.

\section{Experimental setup} \label{experiment}

The experiments were performed in a cylindrical container with inner 
diameter $D=2R=$\unit[180]{mm} and aspect ratio $\Gamma=D/H=1$, where $H$ 
is the height. On top and bottom, the cylinder is bounded by two  
\unit[220]{mm} diameter, \unit[25]{mm} thick uncoated copper plates. 
The sidewalls consist of polyether ether ketone (PEEK). The cell is 
filled with the liquid metal alloy GaInSn with Prandtl number Pr$=0.029$. 
Other relevant physical parameters at $20^{\circ}$\,C are \cite{Plevachuk2015}: 
density $\rho=6350$\,kg/m$^3$, kinematic viscosity 
$\nu=3.44 \times 10^{-7}$\,m$^2$/s, and electrical conductivity 
$\sigma=3.27\times 10^6$ ($\Omega$ m)$^{-1}$. From the latter two
values, we derive a magnetic Prandtl number 
$Pm=\mu_0 \sigma \nu=1.40 \times 10^{-6}$. 

\begin{figure}[t!]
\centering
\includegraphics[width=0.3\textwidth]{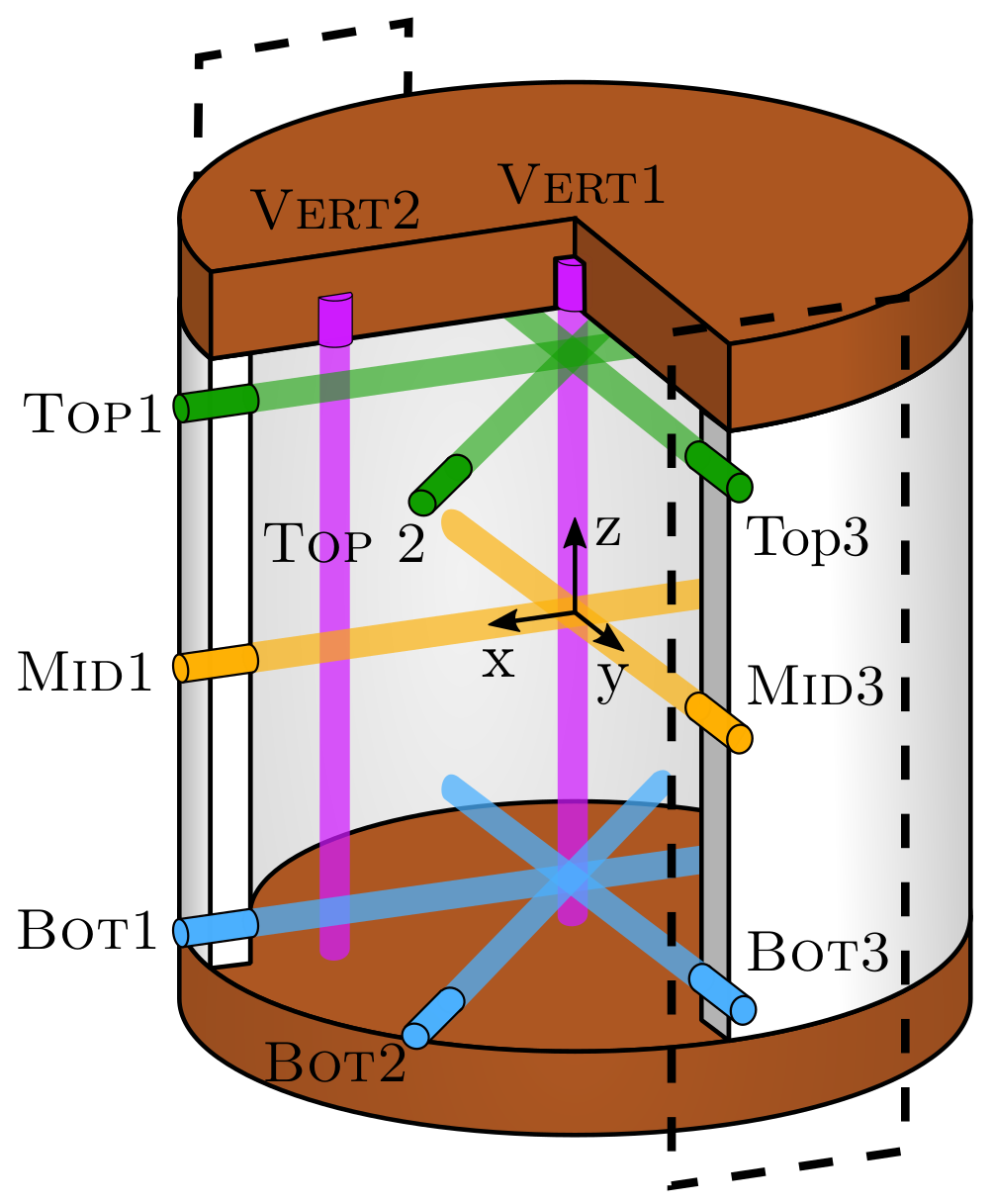}
\caption{Schematic setup of the experimental cell showing the UDV sensor 
placement and the measurement paths. The arrangement of the excitation 
coils lying on the y-axis is indicated by the dashed lines.}
\label{setup}
\end{figure}

The configuration of sensor placements is illustrated in Fig.\,\ref{setup}. 
To measure the radial velocities, eight Ultrasound Doppler Velocimetry 
(UDV) sensors are located at three different heights in the sidewall of 
the cell. They have a flat head, \unit[8]{mm} in diameter and are excited 
with \unit[7.8-8]{MHz} pulses. 
The sampling time for one velocity profile is \mbox{\unit[2.7$\pm$0.2]{s}}, 
if not stated otherwise. Starting from the bottom boundary between 
copper and the liquid alloy, the measurement planes ``Bot'', ``Mid'' and 
``Top'' are located at heights of 10, 90 and \unit[170]{mm}, respectively. 
At each height, two sensors called ``1'' and ``3'' are placed with an 
angle of $\pi/2$ between them. Additionally, the ``Top'' and ``Bot'' 
levels each have an additional sensor ``2'' at the bisection between 
sensors ``1'' and ``3''. Two further UDV sensors for measuring the 
vertical flow component are placed in the 
top copper plate at radial positions $r/R=0$ and $r/R=0.8$. A more 
detailed description of the cell can be found in \cite{Zuerner2019}.

\begin{figure}
	\centering
	\includegraphics[width=0.5\textwidth]{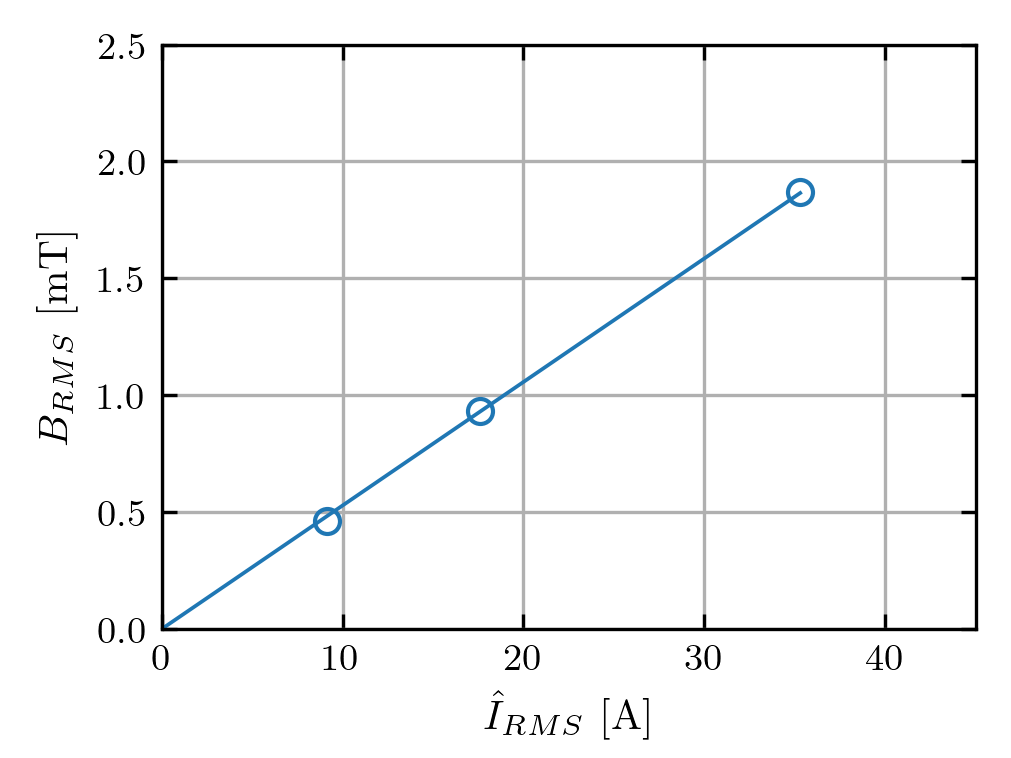}
	\caption{RMS magnetic flux density at \unit[25]{Hz} in the centre 
	of MultiMAG depending 
	on the RMS value of the AC current. The experimental cell was 
	not installed during this 
	measurement.}
	\label{setupMagfield}
\end{figure}

The $m=2$ flow is driven by AC-currents through two rectangular, 
stretched coils which are situated on opposite sides of the cylinder 
as delineated in  Fig.\,\ref{setup}. They consist of 80 turns each 
with a total inner height of \unit[450]{mm} and an average distance 
between the long leg and the x-z  centre plane 
of \unit[145]{mm}\cite{Pal2009}. When a current is run through the 
coils a magnetic field is generated, which is symmetrical to this 
centre plane.  A tunable AC power supply is used to create an 
alternating current with defined amplitude and frequency in the 
coils. The polarity of the coils is assigned in a way that the 
magnetic field is concordant in both solenoids. This 
configuration has turned out advantageous for generating the 
desired flow field\cite{StepanovStefani2019}. For the case that 
the vessel is not installed, the correlation of the measured flux 
density and the applied coil current is depicted in 
Fig.\,\ref{setupMagfield}.

\begin{figure}
	\centering
	\includegraphics[width=0.5\textwidth]{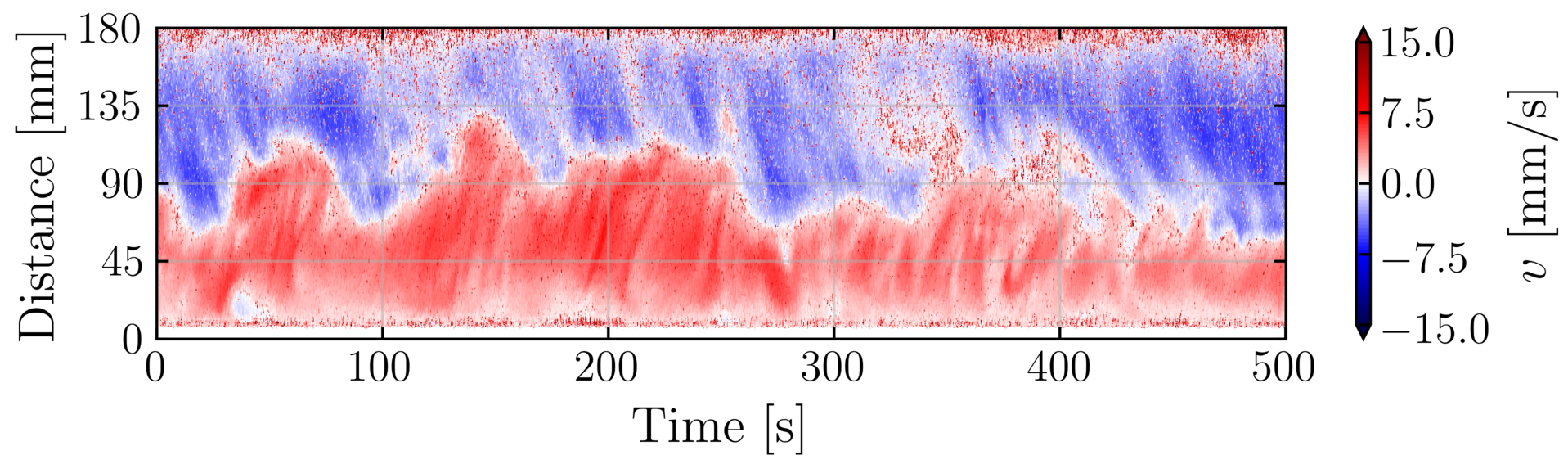}
	\caption{Contour-plot of {\sc Mid1} sensor data 
	at \unit[9]{A} and \unit[25]{Hz}. The 
	sampling time of this particular measurement was 
	reduced to \unit[75]{ms}. The unstable behaviour and the typical 
	timescale of the flow changes are apparent.}
	\label{setupChanOverview}
\end{figure}

For the particular choice of an AC-current with RMS value \unit[9]{A} 
and a frequency of \unit[25]{Hz},
Fig.\,\ref{setupChanOverview} shows an exemplary measurement
result of sensor {\sc Mid1}, i.e. the sensor at mid-height with beamline 
in $x$-direction which is perpendicular to the 
$y$-axis (connecting the two excitation coils). 
In principle, this measurement shows a flow that is directed away from
the walls, leading to a positive projection (red) onto the ultrasound beam 
for lower distances, and  to a negative projection (blue) for higher
distances from the UDV sensor. Not surprisingly, the resulting
stagnation point is rather unstable, leading to a ``restless'' 
behaviour of the flow.

\begin{figure}
	\centering
	\includegraphics[width=0.5\textwidth]{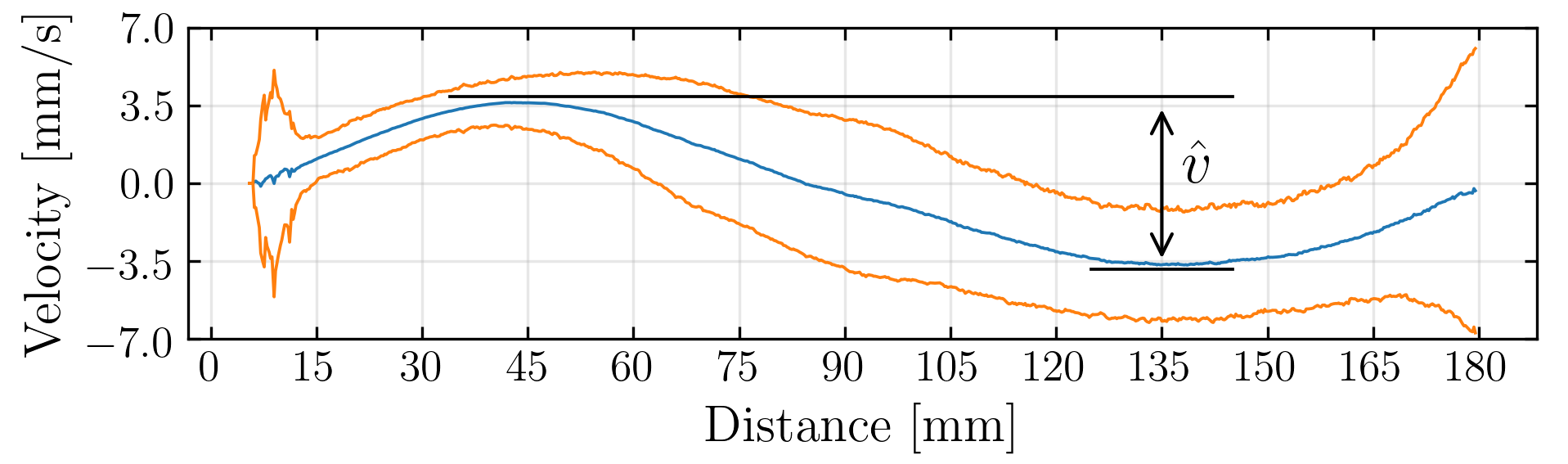}
	\caption{Time averaged velocity (blue) and standard 
	deviation (orange) of the data of figure \ref{setupChanOverview}. 
	The velocity value taken for comparison of parameters is 
	the peak to peak amplitude of the averaged velocity $\hat{v}$.}
	\label{setupMeanvelocity}
\end{figure}

Nevertheless, we can define a time average for all the profiles
measured during this run, which 
yields a sinusoidal dependence on the depth of the ultrasound beam 
(Fig.\,\ref{setupMeanvelocity}), with a peak-to-peak
amplitude $\hat{v}\sim 7$\;mm/s in this specific case.
Every measurement encompasses between 1300 and 1500 profiles  
which, in turn, 
are averages over 100 to 250 UDV emissions (typically 150). 

\section{Numerical simulations} \label{numerics}

\begin{figure}[tbp]
	\centering
	\includegraphics[width=0.4\textwidth]{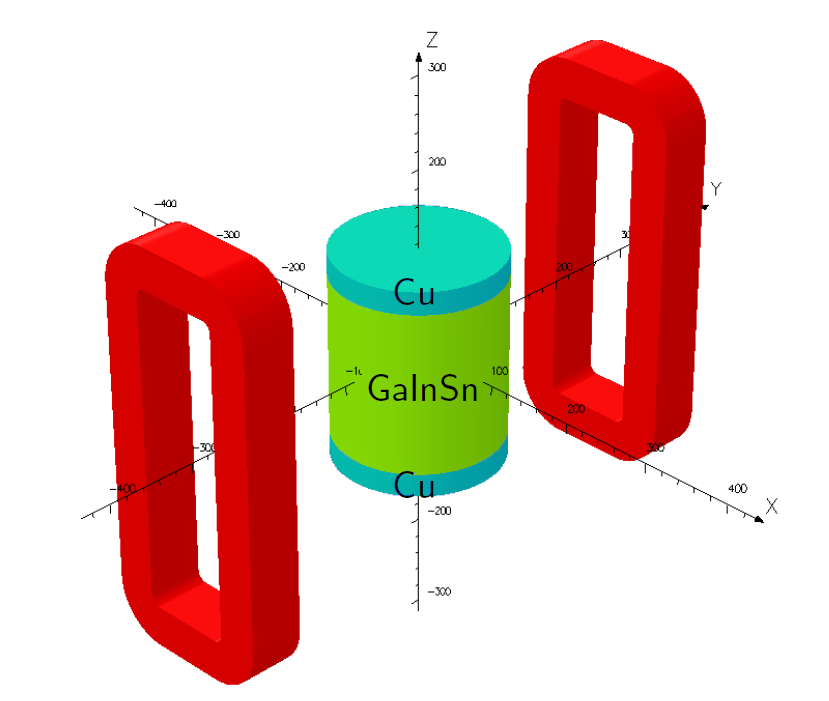}
	\caption{Model setup for pre-computing the electromagnetic 
	force density in Opera 1.7. The two coils of the MultiMAG system
	which are used in our experiment are indicated in red.}
	\label{opera}
\end{figure}

The corresponding numerical simulations of the tide-like liquid metal flow 
were performed using the open source code library 
OpenFOAM$^\copyright$ 5.x \cite{openfoam}. 
The geometry of the entire model (Fig.\,\ref{opera})
emulates as close as possible the real experiment, 
including all physical parameters. The flow in the cell was 
computed solving the incompressible Navier-Stokes equation
\begin{equation} \label{ns}
\frac{\partial \bi u}{\partial t} + (\bi u \cdot \nabla ) \bi u
= - \nabla p + \nu \nabla^2 \bi u + \bi f_{EM} \, ,
\end{equation}
where the time-constant electromagnetic force density $\bi f_{EM}$ 
was pre-computed with 
Opera 1.7. Opera uses the FEA-Method to solve the 
Maxwell equations and to output the body force \cite{opera2018}. 
At all solid 
container walls the no-slip condition $u=0$ was chosen as boundary 
condition for the 
flow field. No turbulence model was used in the simulation of the 
fluid flow.

\begin{figure}[tbp]
	\centering
	\includegraphics[width=0.5\textwidth]{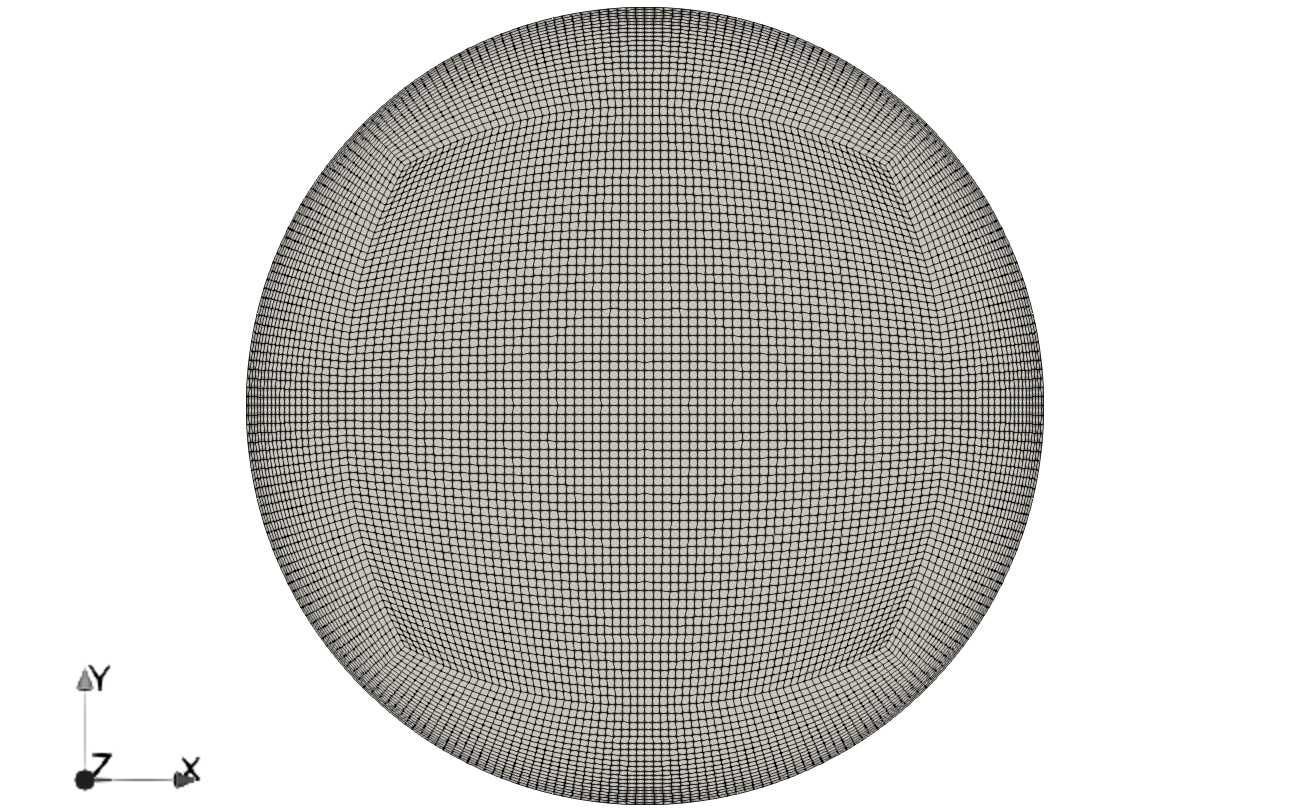}
	\caption{Symmetric hexahedral mesh with 1.3 million elements, cut in the 
	central $x-y$ plane, as generated for the OpenFOAM simulations.}
	\label{mesh1}
\end{figure}

Anticipating the typical Reynolds number of the generated tide-like flow, 
the cylindrical volume of \unit[0.00458] {m$^3$} of the experimental cell 
was discretised using 1.3 million hexahedral cells. The smallest cell 
of the plane symmetric mesh has a volume of \unit[0.373] {mm$^3$} and the biggest 
one \unit[7.89] {mm$^3$} (see Fig.\,\ref{mesh1} for an impression). 
Close to the walls, the cells are contracted.

\begin{figure}[tbp]
	\centering
	\includegraphics[width=0.33\textwidth]{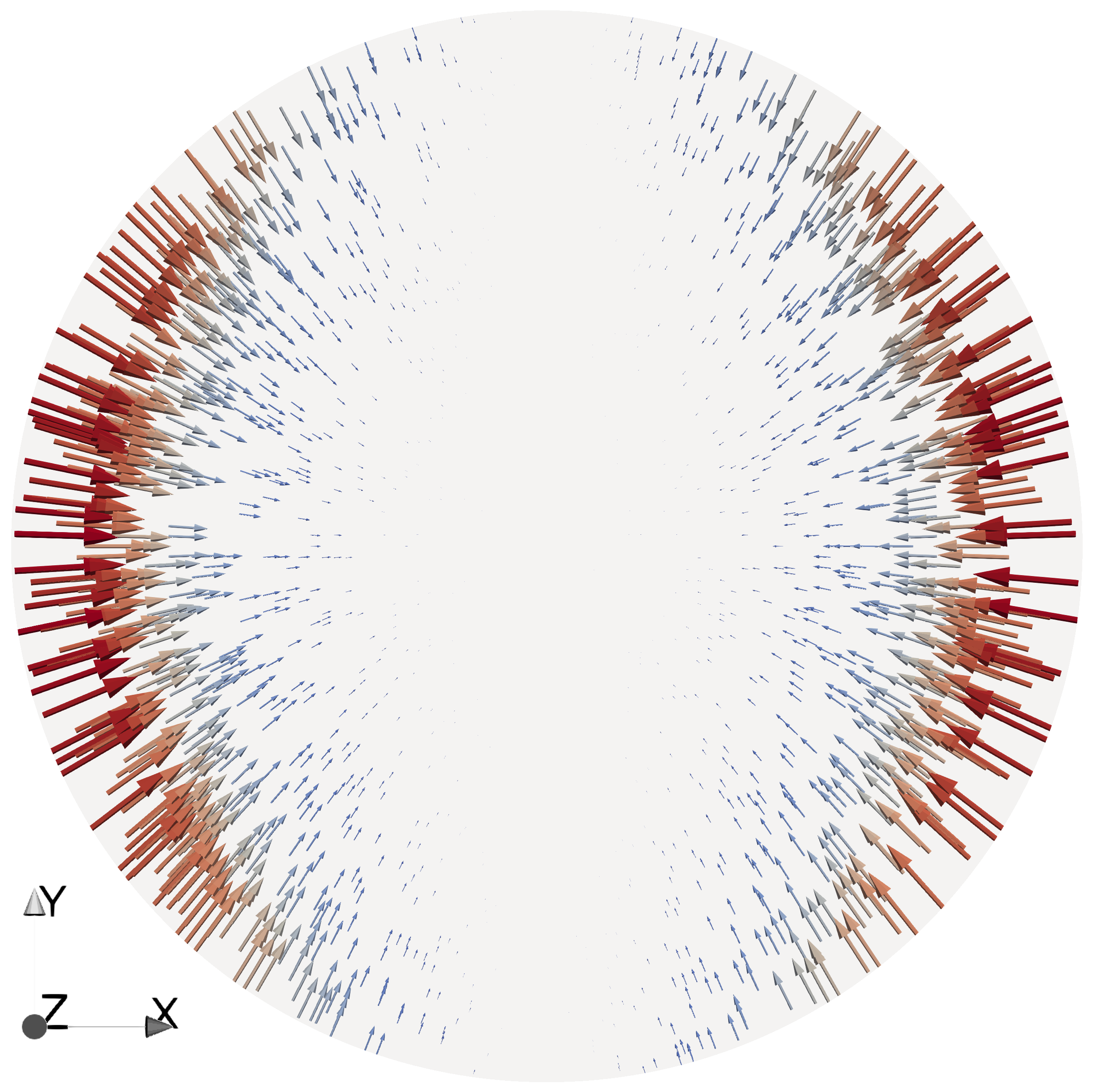}
	\caption{Structure of the body force as computed with 
	Opera  
	and imported into OpenFOAM. The force points mainly radially inward 
	along the $x$-axis.}
	\label{Screen-BF}
\end{figure}

For the correct computation of the electromagnetic body force it 
is important to take into account the copper plates at the top and
the bottom which tend to homogenize (in vertical direction) 
the body force in the fluid region.
With increasing frequency this force is concentrated at the side walls, 
due to the skin effect. A typical force
structure is shown (for the special case \unit[9.7]{A} and \unit[25]{Hz})
in Fig.\,\ref{Screen-BF}. Evidently, the force mainly pushes radially inward
along the $x$-axis, generating the two oppositely directed jets which 
were already
visible in the experimental results (Fig.\,\ref{setupChanOverview}).

\begin{figure}[tbp]
	\centering
	\includegraphics[width=0.5\textwidth]{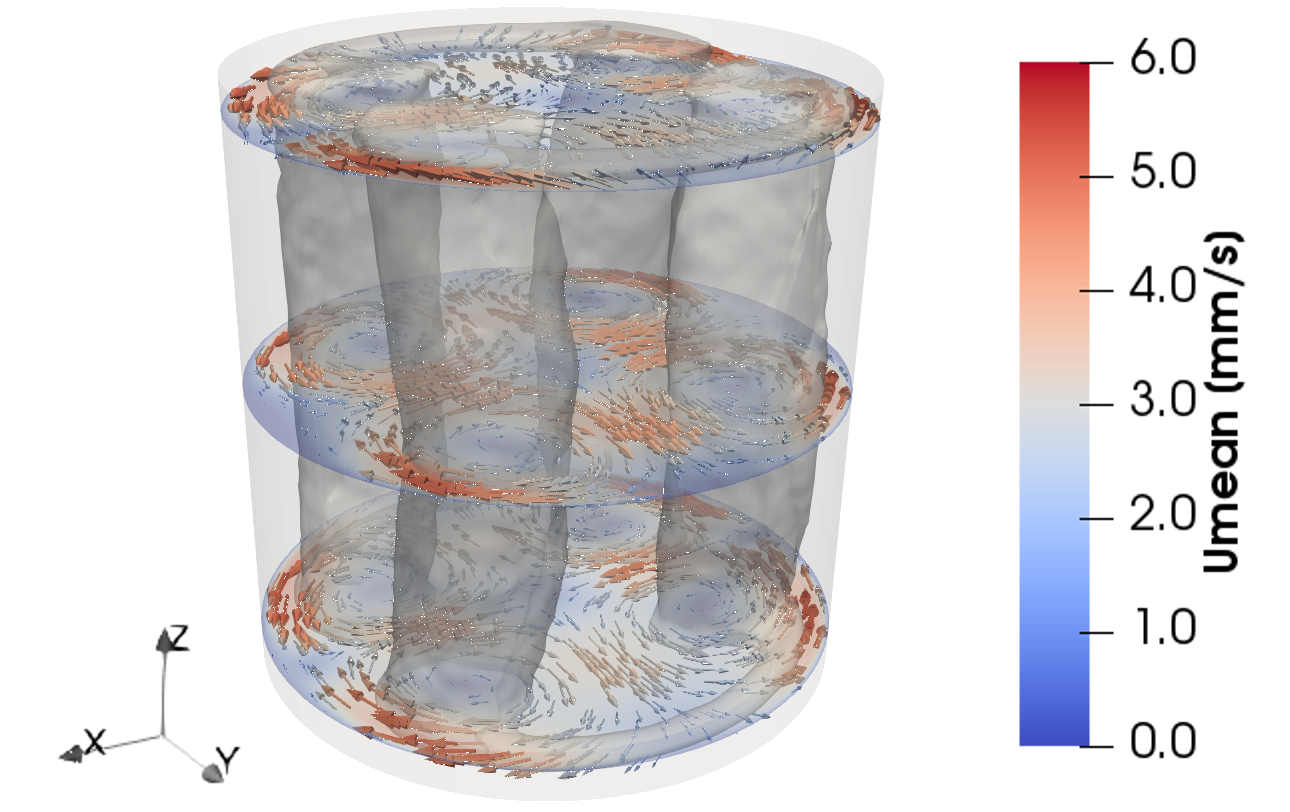}
	\caption{Velocity field time averaged over \unit[10000]{s}, 
	showing four quasi-two-dimensional vortices. 
	I=\unit[9.8]{A} and f=\unit[25]{Hz}}
	\label{CFDRollesUmean}
\end{figure}

\begin{figure}[tbp]
	\centering
	\includegraphics[width=0.5\textwidth]{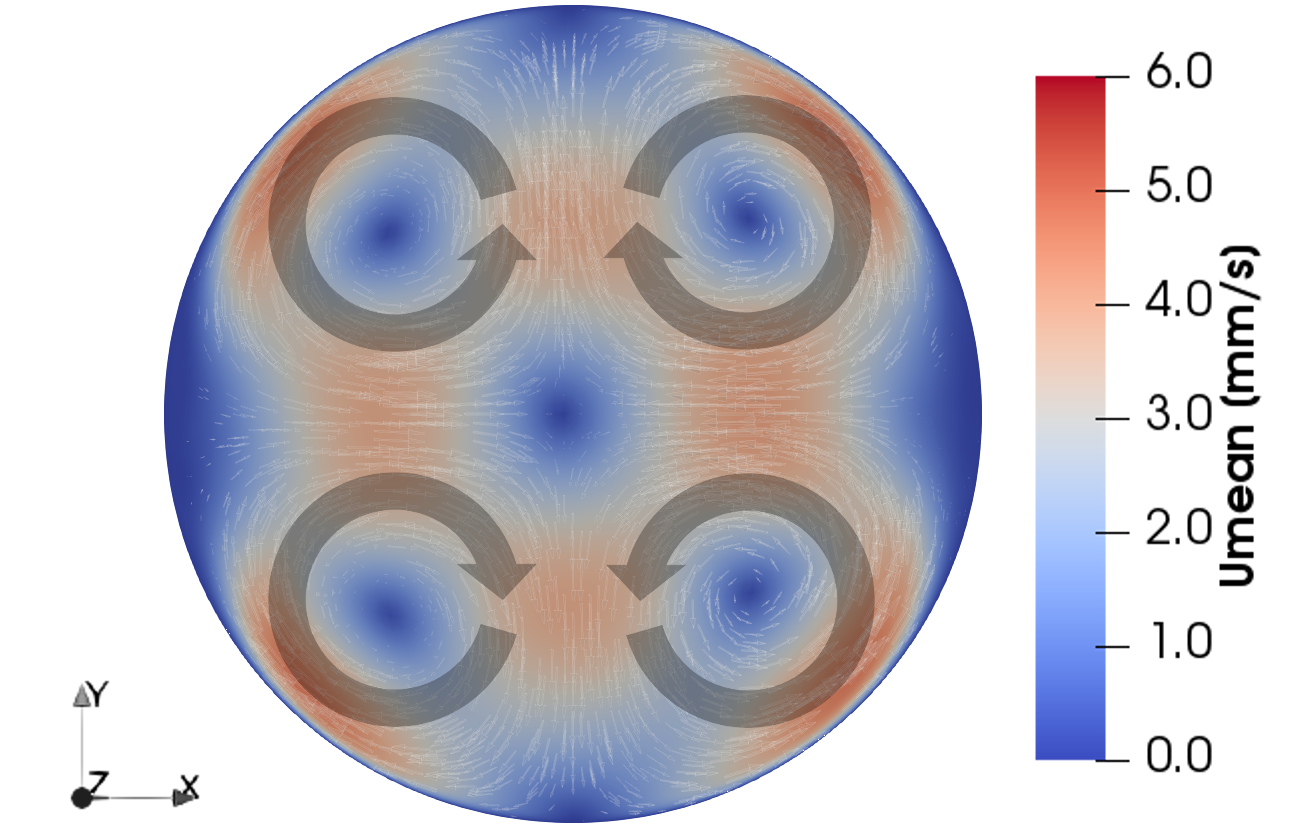}
	\caption{Velocity field time averaged over \unit[10000]{s}, 
	highlighting  the four quasi-two-dimensional vortices at 
	mid-height. I=\unit[9.8]{A} and f=\unit[25]{Hz}}
	\label{CFDUmean}
\end{figure}

\begin{figure}[tbp]
	\centering
	\includegraphics[width=0.5\textwidth]{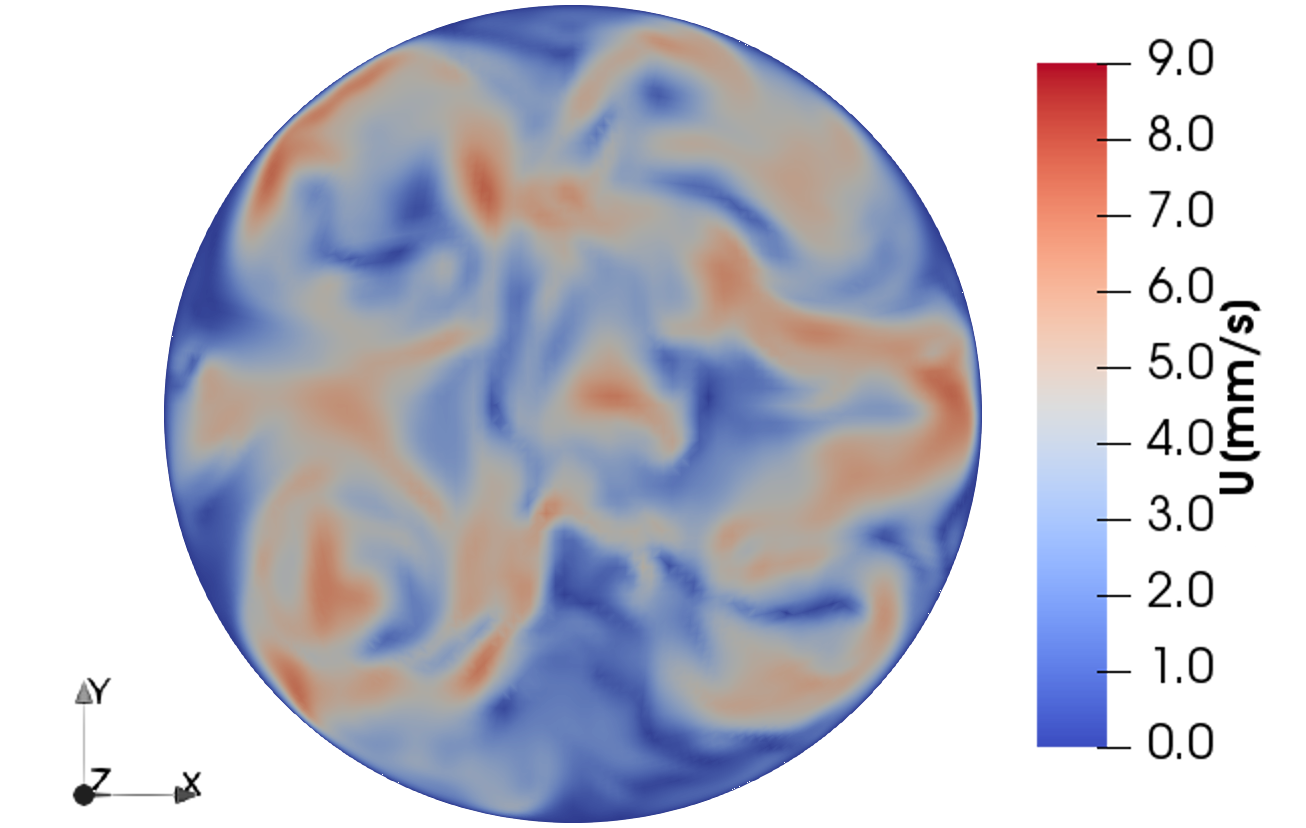}
	\caption{Instantaneous flow structure  at mid-height and t=\unit[2760]{s}. 
	I=\unit[9.8]{A} and f=\unit[25]{Hz}}
	\label{CFDU2760s}
\end{figure}

In the OpenFOAM simulation, at t = \unit[0]{s} the body force is applied 
to a u = \unit[0]{m/s} base state. After some time, the flow reaches 
a quasi-steady state. The specific flow structure, produced by an 
excitation current with 
\unit[2.4]{A} and \unit[25]{Hz} and averaged over a simulation time of 
10000 s, is illustrated in Fig.\,\ref{CFDRollesUmean}.
At all three selected heights, it shows the presence of four very similar,
quasi-two-dimensional vortices, which are driven by the radially inward 
directed body force along the $x$-axis.
Figure\,\ref{CFDUmean} depicts in more detail this vortex structure 
in the mid-height $x-y$-plane. While the time-averaged flow field shows
a very regular structure, the instantaneous flow can significantly deviate
from that average. This is illustrated in Fig.\,\ref{CFDU2760s} in which
the four-vortex structure is barely visible anymore.

\section{Results} \label{results}

In this section we will present and compare the 
experimentally measured velocity fields with the numerically
determined ones. In most detail we will discuss the results
for the three current amplitudes  
\unit[9.7]{A}, \unit[5.9]{A}, \unit[2.4]{A},  
with frequency \unit[25]{Hz}, while a few more results will be 
presented for currents up to \unit[50]{A} and frequencies 
up to \unit[200]{Hz}.

The parameter combination \unit[9.7]{A} and
\unit[25]{Hz} produces relatively high flow velocities 
which can be clearly identified by UDV. For six selected sensors,
Fig.\,\ref{resultsContour9_7A} shows the contour plots of the
actually measured signal (left column) and of corresponding 
``virtual sensors''
as extracted from the numerical simulation (right column). 
Evidently, sensor ``Mid3'', 
measuring along the $y$-axis at mid-height, shows a stable flow
structure which points from the center towards the wall, leading
to a negative projection (blue) close to the sensor, and a 
positive projection
(red) at greater depth of the ultrasonic beam. While the division 
point between
these two outward directed jets remains rather stable, there are 
some fluctuations 
both in the measured amplitudes and in the simulation.

In comparison with this quiet behaviour, sensor ``Mid1'' shows much stronger 
fluctuations. As mentioned above, these fluctuations are due to an
instability of the stagnation point where the two inward directed jets 
converge toward each other. This leads to a bi-modal behaviour, 
with a tendency of the 
stagnation point to stay for most of the time at either side of the center
rather than at the center itself. 

The behaviour at the top and the bottom is not very different from that 
at mid-height, as exemplified here for the inward directed jets at the top
(``Top1'') and the outward directed jets at the bottom (``Bot3'').

The most unstable situation is found at the sensors of position ``2" 
with a 45 degree angle to the coils which are situated between 
sensors ``1'' and ``3''. Evidently, at sensor ``Top2'' the flow 
projection onto the ultrasound beam fluctuates strongly and correlates 
thereby significantly with the movement of the stagnation point 
of ``Top1''. ``Top2'' is also sensitive to azimuthal rotations 
of the flow structure about the cylinder axis. A similar fluctuation 
behaviour is seen in ``Bot2''.

\begin{widetext}
\onecolumngrid

\begin{figure}
	\centering
	\includegraphics[width=\textwidth]{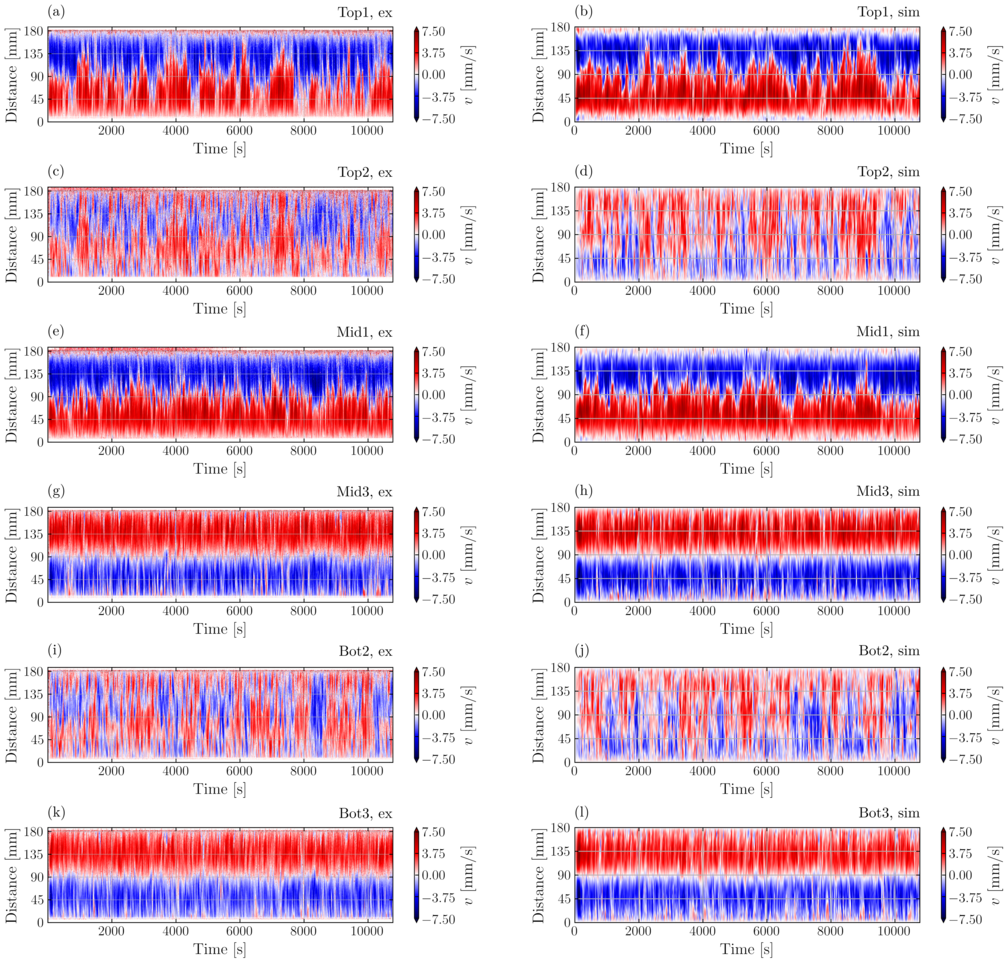}
	
	\caption{Contour plots of the flow speed for \unit[9.7]{A} 
	and \unit[25]{Hz}. 
	The left column shows measurement data, while the right 
	column shows virtual sensor 
	data from the simulations.\label{resultsContour9_7A}}
\end{figure}
\twocolumngrid
\end{widetext}

The corresponding contour plots for the lower current amplitudes
\unit[5.9]{A} and \unit[2.45]{A}  are presented in Fig.\,\ref{resultsContour5_9A} and
Fig.\,\ref{resultsContour2_45A}, respectively.
Compared with the previous case of \unit[9.7]{A}, both the flow amplitude 
and 
the typical frequency of the oscillation of the stagnation point
decreases for the case with \unit[5.9]{A}. For the extreme case with
\unit[2.45]{A}, we observe a very long transient behaviour for the
onset of the fluctuations. The typical flow velocities are 
in the order of 1 mm/s which makes their clear identification by UDV 
challenging.

\begin{widetext}
\onecolumngrid

\begin{figure}
	\centering
	\includegraphics[width=\textwidth]{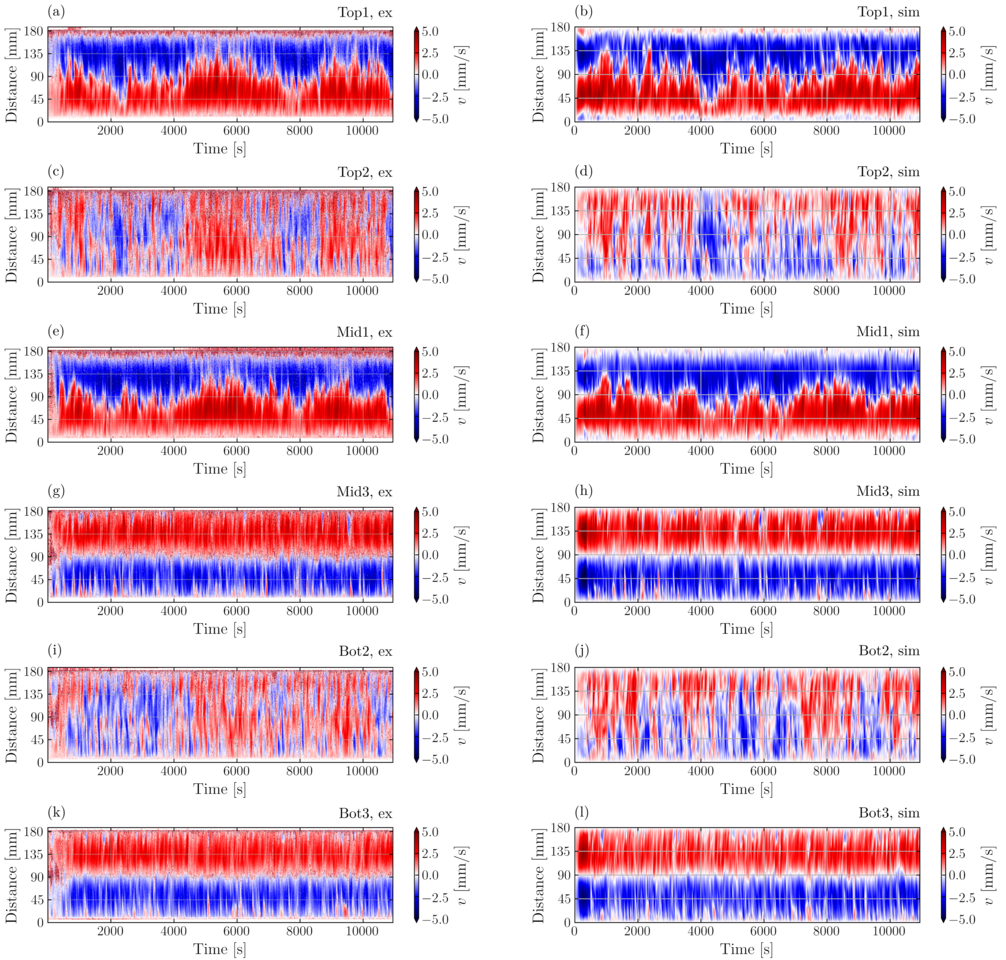}
	
	\caption{Contour plots of the flow speed for \unit[5.9]{A} 
	and \unit[25]{Hz}. 
	The left column shows measurement data, while the right 
	column shows virtual sensor 
	data from the simulations.\label{resultsContour5_9A}}
\end{figure}
\twocolumngrid
\end{widetext}

\begin{widetext}
\onecolumngrid

\begin{figure}
	\centering
	\includegraphics[width=\textwidth]{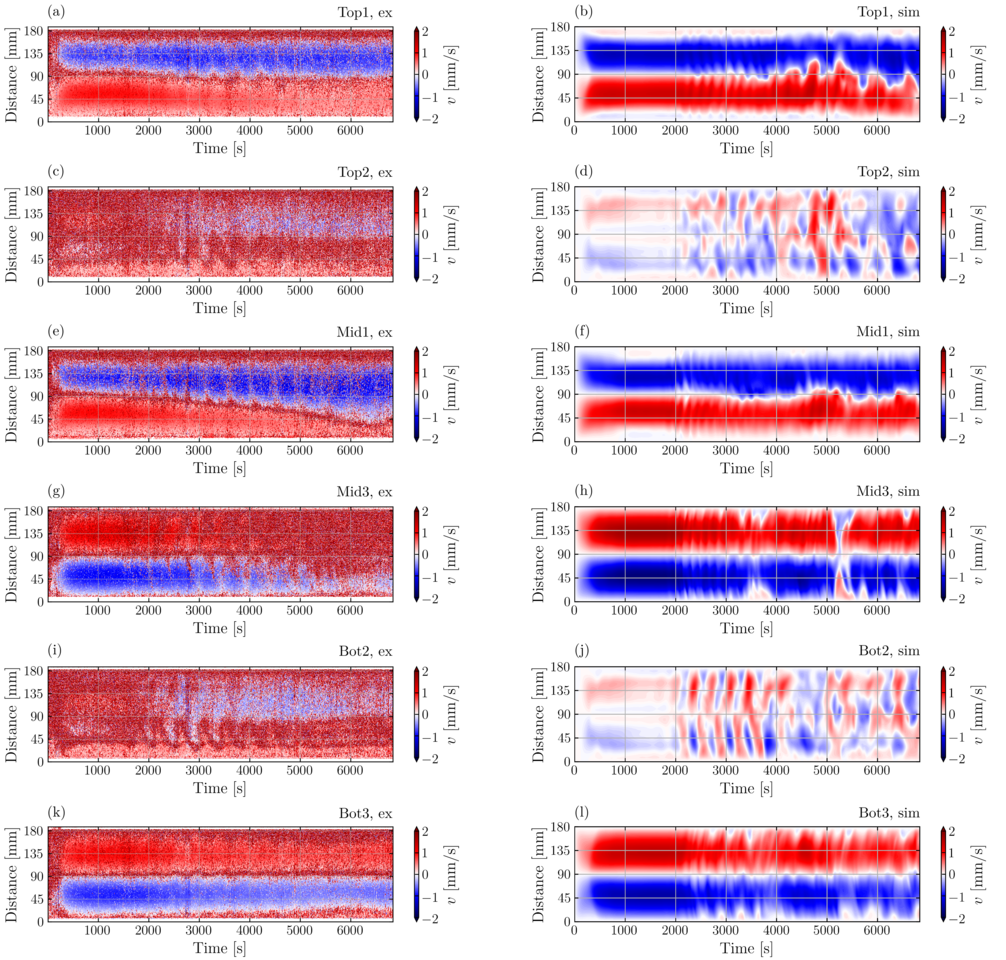}
	
	\caption{Contour plots of the flow speed for \unit[2.45]{A} 
	and \unit[25]{Hz}. 
	The left column shows measurement data, while the right 
	column shows virtual sensor 
	data from the simulations.\label{resultsContour2_45A}}
\end{figure}
\twocolumngrid
\end{widetext}

\begin{figure}
	\centering
	\includegraphics[width=0.5\textwidth]{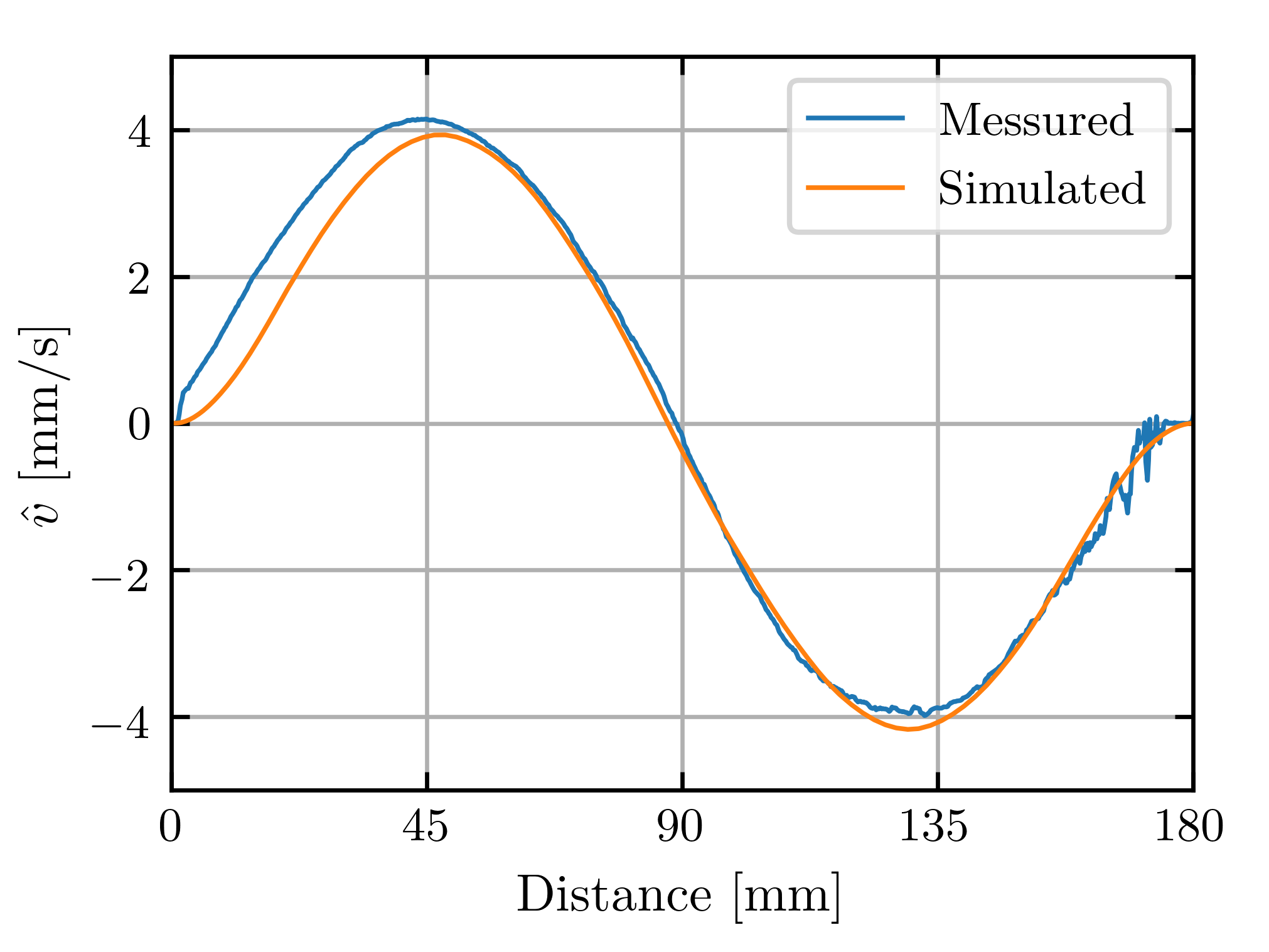}
	\caption{Comparison of simulated and measured average flow speed for sensor {\sc Mid1}. I=\unit[9.8]{A}, f=\unit[25]{Hz}}
	\label{resultsComparisonCFDExpMid1}
\end{figure}

\begin{figure}
	\centering
	\includegraphics[width=0.5\textwidth]{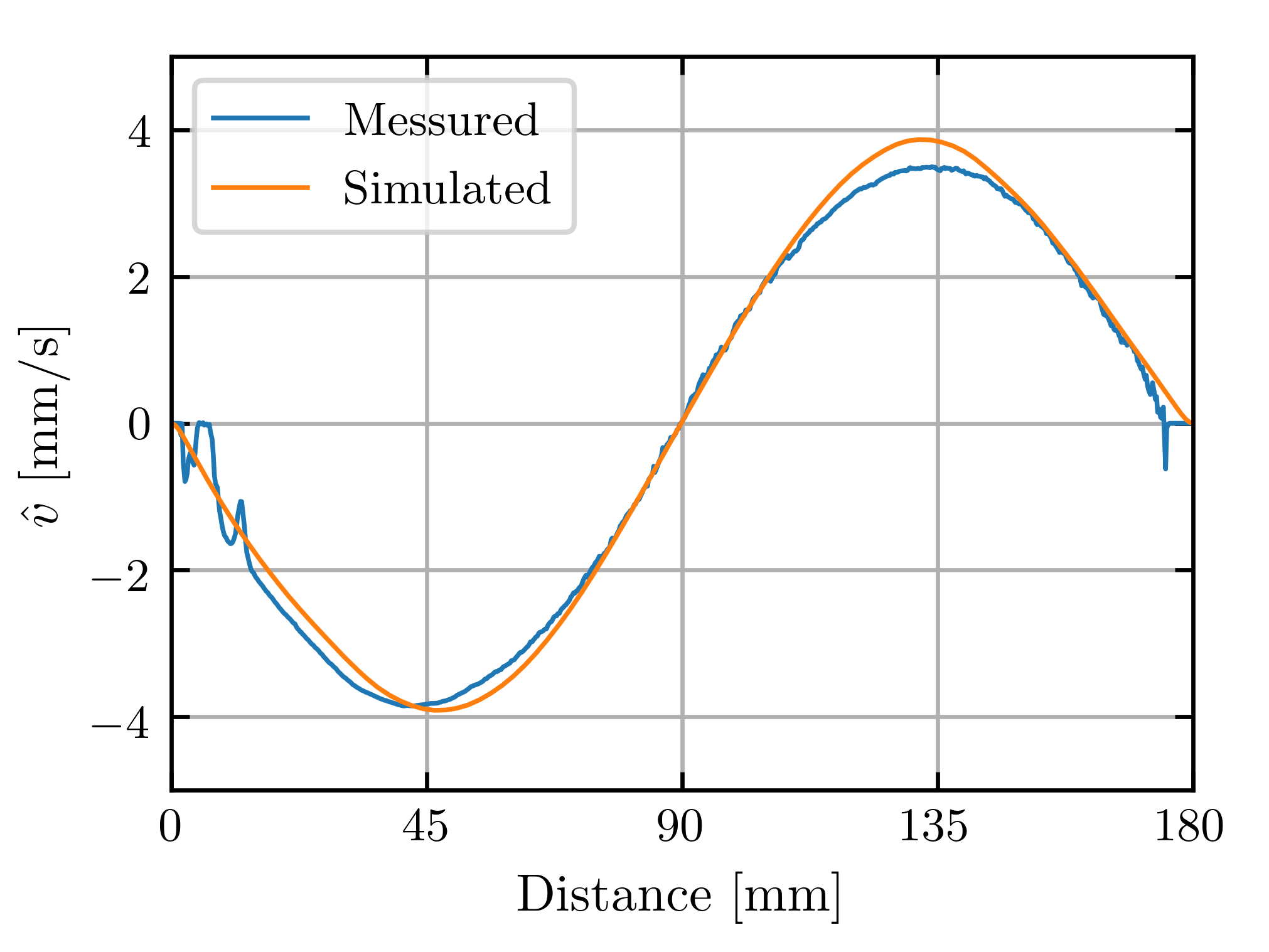}
	\caption{Comparison of simulated and measured average flow speed for sensor {\sc Mid3}. I=\unit[9.8]{A},
	  f=\unit[25]{Hz}}
	\label{resultsComparisonCFDExpMid3}
\end{figure}

Apart from the mentioned oscillatory motion of the stagnation point, 
for the time-averaged flow structure 
we find a nearly perfect agreement between measurement and simulation. 
For the case with \unit[9.7]{A} and \unit[25]{Hz}, this is illustrated 
for sensors Mid1 and Mid3 in Fig.\,\ref{resultsComparisonCFDExpMid1} 
and Fig.\,\ref{resultsComparisonCFDExpMid3}, respectively.


The bi-modal behaviour of the stagnation point of the two inward directed 
jets is analysed in Fig.\,\ref{resultsHistograms}. It shows, for the three 
considered current amplitudes, the experimentally measured and numerically 
determined probability densities of the velocity in $x$-direction (as 
seen by sensor Mid1) in the centre of the cell. Apart from some slight 
quantitative differences, experiments and simulations show a distinct 
bi-modality of this velocity component. Due to the finite 
experimental/simulation time and possibly some geometric imperfections 
of the experiment, the bi-modality is not perfectly symmetric. For the 
lowest current  \unit[2.45]{A} the period of the stagnation point 
oscillation is already so large that the stagnation point stays 
completely on one side.

\begin{widetext}
\onecolumngrid

\begin{figure}
	\centering
	\includegraphics[width=\textwidth]{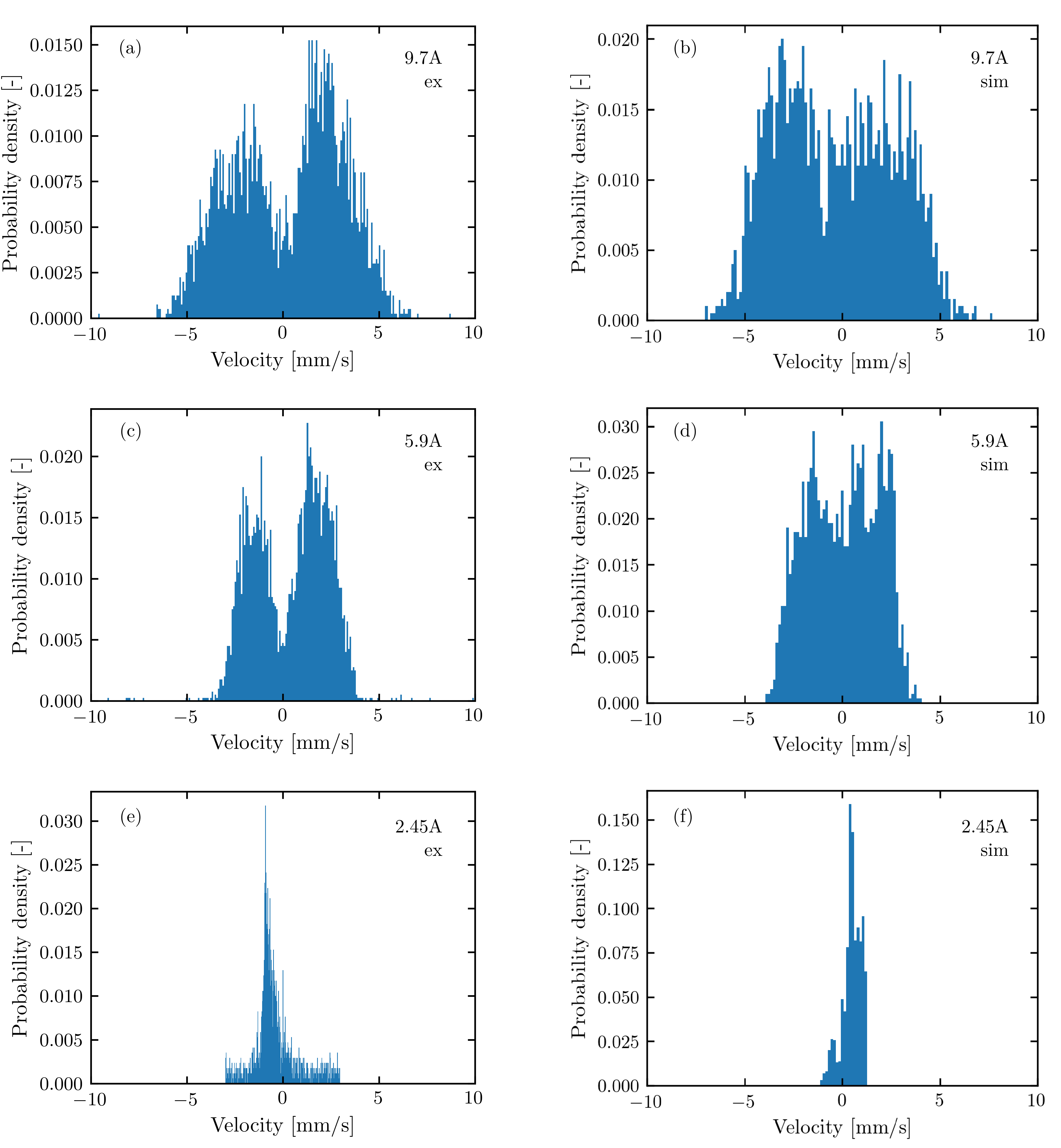}
	\caption{Histograms of the velocity of sensor {\sc Mid1} 
	at the centre of the cylinder. On the left experimental data, 
	on the right, CFD velocities.}
	\label{resultsHistograms}
\end{figure}
\twocolumngrid
\end{widetext}

\begin{figure}
	\centering
	\includegraphics[width=0.5\textwidth]{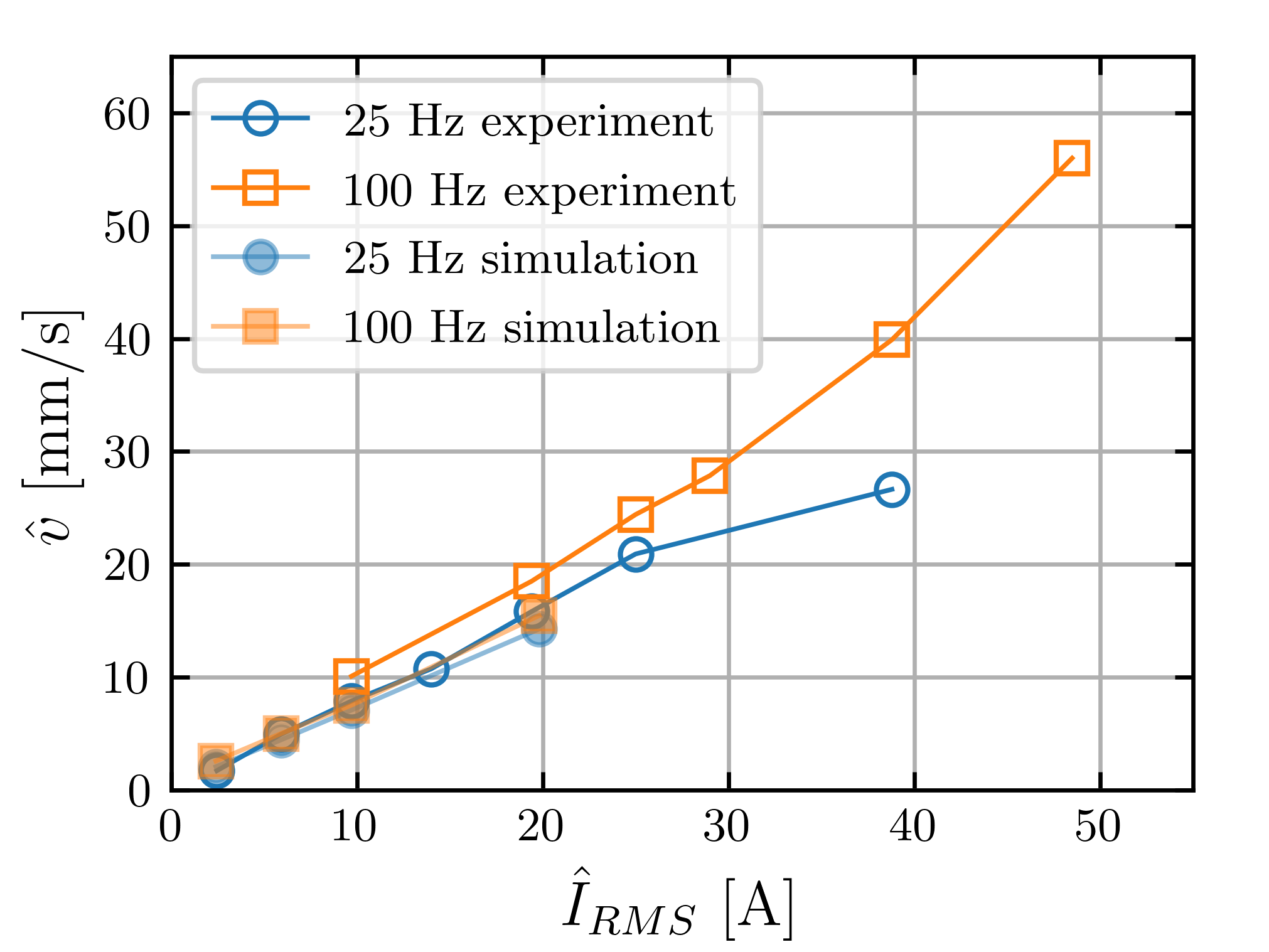}
	\caption{Current variation. X-axis: the root-mean-square 
	of the alternating current in the RMF-coils. Y-axis: 
	peak-to-peak velocity as described in Fig.\,\ref{setupMeanvelocity}. 
	The curves differ by the frequency of the alternating 
	current. The 2.45A experimental point is taken from a 
	30 minutes measurement, which is comparably short to 
	the fluid motion time scale.}
	\label{resultStromvar}
\end{figure}

After focusing on the flow behaviour for three particular, and 
weaker, current amplitudes, we will now summarize the dependence 
of the averaged flow amplitudes on the amplitude and the frequency 
of the current. For frequencies \unit[25]{Hz} and \unit[100]{Hz}, 
Fig.\,\ref{resultStromvar} indicates a mainly linear dependence 
of the flow amplitude on the current amplitude, with some notable 
deviations above \unit[30]{A}. Corresponding simulations, 
carried out for \unit[25]{Hz} until \unit[30]{A}, show in 
general a good agreement with the experiments.

\begin{figure}
	\centering
	\includegraphics[width=0.5\textwidth]{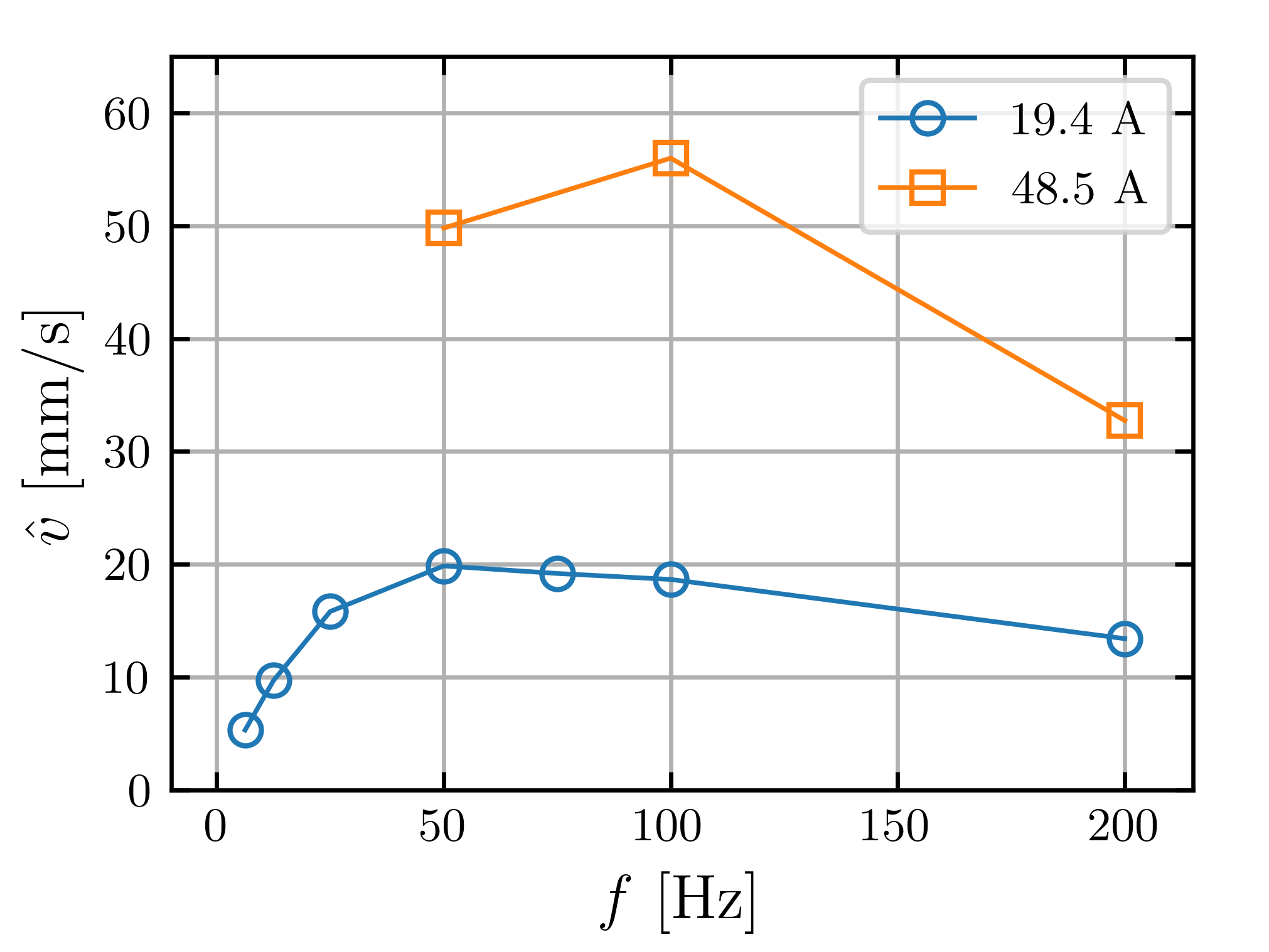}
	\caption{Frequency variation. X-axis: frequency of the 
	alternating current in the RMF-coils. Y-axis: 
	peak-to-peak velocity as described in 
	Fig.\,\ref{setupMeanvelocity}. The curves are 
	dependent on the current amplitude.}	
	\label{resultFrequenzvar}
\end{figure}

For two exemplary current amplitudes \unit[19.4]{A} and \unit[48.5]{A},
Fig.\,\ref{resultFrequenzvar} illustrates the dependence of the 
flow amplitude on the frequency of the current. Here we observe 
a maximum approximately at \unit[50-100]{Hz}, depending mildly 
on the current amplitude. At too low frequencies, the slowly 
changing magnetic field only weakly couples to the flow, while 
for too high frequencies the field is hindered in penetrating 
the vessel by the skin effect. The frequency \unit[25]{Hz} 
seems particularly suitable for the later synchronization 
experiments since on the one hand, it is only slightly 
below the optimum frequency, and has on the other hand a 
large enough penetration depth to produce a rather smooth 
forcing distribution that is not too closely concentrated 
at the walls.\\


\begin{figure}
	\centering
	\includegraphics[width=0.48\textwidth]{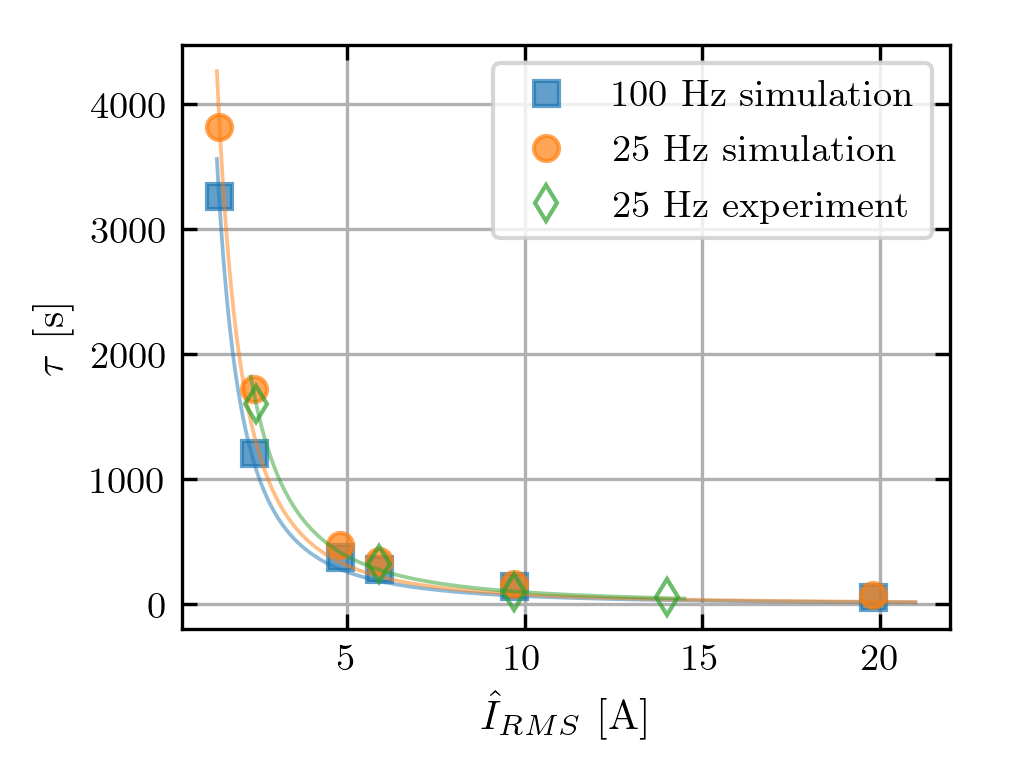}
	\caption{Time until the saturated value of $\bar{u^2}$ is 
	reached. The experimental points are an estimate from 
	the destabilization of the stagnation point in 
	the {\sc Mid1} data. Fit parameters are listed 
	in Table \ref{tab.anlaufParameter}.}
	\label{resultsAnlauf}
\end{figure}

Our last result concerns the dependence of some typical time-scales 
on the current amplitude. While neither experiments nor simulations 
are long enough to allow for a reasonable statistics of the periods 
of the oscillation between different stagnation points, we can still 
determine the build-up time for the saturated mean square 
velocity $\bar{u^2}$, where typically the oscillation starts. 
The corresponding dependencies, obtained numerically and 
experimentally, are plotted in Fig.\,\ref{resultsAnlauf}. 
As $\bar{u^2}$ is not available in the experiments, the points 
were estimated from the destabilization of the stagnation point 
in the contour plots.

According to the Navier-Stokes equation and the induction 
equation, the acceleration of the fluid is proportional to 
the squared current amplitude. Therefore the build-up time 
$\tau_b$ is modelled to decrease with increasing current 
amplitude $\hat{I}_{RMS}$ following the equation
\begin{equation}
\tau_b = \frac{1}{a \cdot \hat{I}_{RMS}^2} \label{eqn.AnlaufFit}
\end{equation}
The determined fit parameters are listed in Table \ref{tab.anlaufParameter}.
\begin{table}
	\caption{Fit parameters ``a'' for Eq.\,\ref{eqn.AnlaufFit} 
	used in Fig.\,\ref{resultsAnlauf}, with $r^2$ denoting 
	the square of Pearsons correlation coefficient.} 
	\begin{tabular}{rcc}
		\hline
		& a $[s^{-1} A^{-2}]$ & $r^2$\\
		\unit[25]{Hz} sim & 1.286e-4 & 0.9812\\ 
		\unit[25]{Hz} exp & 1.037e-4 & 0.9987 \\ 
		\unit[100]{Hz} sim & 1.541e-4  & 0.9956 \\ 
		\hline
	\end{tabular}
	\label{tab.anlaufParameter}
\end{table}

At the lowest applied current of \unit[2.45]{A} with a frequency 
of \unit[25]{Hz}, the flow becomes unstable at 
$\tau_b\approx\unit[1800]{s}$. With view on the later RBC 
experiment \cite{LSC-Till-2019} this exceeds the 
Large-Scale-Circulation (LSC) time by more than one order 
of magnitude. If we consider a sinusoidal swelling 
excitation of the tide-like forcing, the choice of I=\unit[2.45]{A} 
seems therefore not ideal. For this purpose, currents of \unit[5.9]{A} 
with a build-up time $\tau_b\approx\unit[300]{s}$, or  I=\unit[9.7]{A} 
with $\tau_b\approx\unit[150]{s}$, seem to be more suitable. 
However, as the $m=2$ forcing is supposed only to synchronize
the sloshing instability of the LSC and not to dominate the 
entire flow structure in the vessel, for each Rayleigh number 
a optimization of the current will be required.

\section{Conclusions and prospects}\label{conclusions}

In this paper we have compared experimental and numerical results on the
tide-like flow in a cylindrical volume of a liquid metal that 
is driven by AC currents in two oppositely situated coils. 
With Ampere-turns in the order of \unit[1000]{A}
flow speeds of \unit[10]{mm/s} are easily obtainable, despite the 
relatively large distance of the coils from the rim of the liquid metal. 
The dependence of the flow intensity on the AC current frequency shows 
a rather broad plateau between 25 and 100 Hz; at \unit[25]{Hz} we find a 
good compromise between a not too shallow skin depth and a reasonable 
forcing.

In the next step we plan to act with the modulated electromagnetic forcing onto the 
typical LSC flow of RB convection. As shown in \cite{Zuerner2019}, the free 
fall time $\tau_{ff}$ of the LSC  is in the range of a few seconds, 
leading to a typical oscillation frequency of the sloshing  mode of 
$\tau_{osc} \sim 10 \tau_{ff}$ of some tens of seconds. This also correspond 
to the the period of modulating the $m=2$ forcing which we expect will 
lead to a resonant excitation of the sloshing mode, and thereby of the helicity 
oscillation of the LSC.

\section*{Acknowledgments and Data Availability}

This work was supported in frame of the Helmholtz - Russion 
Science Foundation Joint Research Group ''Magnetohydrodynamic 
instabilities'', contract numbers HRSF-0044 and 18-41-06201, 
by the Deutsche Forschungsgemeinschaft with Grant VO 2332/1-1,  
and by the European Research Council (ERC) under the 
European Union's Horizon 2020 research and innovation 
programme (grant agreement No 787544).

The data that support the findings of this study are 
available from the correspo
nding author upon reasonable request.

\end{document}